\documentclass[epjST,final]{svjour}
\usepackage{verbatim}
\usepackage{graphics}
\usepackage{multicol}
\begin{document}
\title{Exploiting magnetic properties of Fe doping in zirconia}
\subtitle{From first-principles simulations to 
the experimental growth and characterization of thin films.}
\author{D. Sangalli\inst{1}\fnmsep\thanks{\email{davide.sangalli@mdm.imm.cnr.it}} \and
        E. Cianci\inst{1}                                                         \and
        A. Lamperti\inst{1}                                                       \and
        R. Ciprian\inst{1}                                                        \and
        F. Albertini\inst{2}                                                      \and
        F. Casoli\inst{2}                                                         \and
        P. Lupo\inst{2}                                                           \and
        L. Nasi\inst{2}                                                           \and
        M. Campanini\inst{2}                                                      \and
        A. Debernardi\inst{1}                                                     }
\institute{Laboratorio MDM - IMM - CNR via C. Olivetti, 2 I-20864 Agrate Brianza (MB) Italy \and
           IMEM-CNR, Parco delle Scienze 37/A – 43124, Parma, Italy}
\abstract{
In this study we explore, both from theoretical and experimental side, the effect of $Fe$ doping in
$ZrO_2$ ($ZrO_2$:$Fe$). By means of first principles simulation we study the magnetization
density and the magnetic interaction between $Fe$ atoms. We also consider how this
is affected by the presence of oxygen vacancies and compare our findings with 
models based on impurity band~\cite{Coey2005} and carrier mediated
magnetic interaction~\cite{Dietl2000}.
Experimentally thin films ($\approx 20 nm$) of $ZrO_2$:$Fe$ at high doping concentration
are grown by atomic layer deposition. We provide experimental evidence that $Fe$ is
uniformly distributed in the $ZrO_2$ by transmission electron microscopy
and energy dispersive X--ray mapping,
while X--ray diffraction evidences the presence of the fluorite crystal structure.
Alternating gradient force magnetometer measurements
show magnetic signal at room temperature, however with low magnetic moment
per atom. Results from experimental measures and theoretical simulations are compared.
}
\maketitle
%
\section{Introduction}
\label{sec:intro}
Dilute magnetic semiconductors (DMS) are materials in which magnetic impurities are introduced
in order to produce a magnetic ground state. These systems have received great attention in
recent years,
since the discovery of carrier induced ferro--magnetism in $(In,Mn)As$~\cite{Ohno1992} and
$(Ga,Mn)As$~\cite{Ohno1996}, and are believed to be fundamental to fabricate spin--based electronic devices.
Recently a new class of DMS has been investigated, namely DMS based on
high-$k$ oxides, i.e. dilute magnetic oxides (DMO),
after the experimental evidence of room temperature magnetism in transition metals (TMs) doped
zirconia~\cite{Hong2012,Coey2005_PRB,Sahoo2009,Kriventsova2006} ($ZrO_2$),
hafnia~\cite{Hong2005,Hong2006} ($HfO_2$),
and titania~\cite{Matsumoto2001,Wang2003} ($TiO_2$)
and the theoretical prediction of high $T_c$ in TMs doped $ZrO_2$~\cite{Ostanin2007,Archer2007}.

The understanding of DMS/DMO physical properties constitutes a challenge for the theory
as the fundamental mechanism leading to ferromagnetic (FM) interaction between the dopants
cannot be explained in terms of simple exchange mechanisms, at least at low doping concentration,
being the latter often too short--ranged.
Among the other, two theoretical models have been proposed to describe FM effects:
the first is based on the presence of impurity states in the crystal, impurity band model (IBM)~\cite{Coey2005},
the other on carriers in spin polarized bands, carrier mediated model (CMM)~\cite{Dietl2000}
which is indeed a refined version of the Zener model.

From the experimental side the inclusion and influence of magnetic dopant, such as
$Fe$, $Co$, $Ni$ and $Mn$, is not clearly understood.
Indeed, while several DMS/DMO have been predicted to have a Curie temperature ($T_c$) above room temperature,
no experimental report of $T_c>300 K$ has been left unchallenged by other studies~\cite{Sato2010}.
Moreover some results suggest that magnetic impurities act as paramagnetic (PM) centers with unusually
long relaxation time~\cite{Gunnlaugsson2010}, at least at very low doping concentrations.

In this manuscript we study iron doped zirconia ($ZrO_2$:$Fe$) focusing our attention
on the magnetic properties of the system.
In Sec.~\ref{sec:structural_properties} we provide a structural characterization
of thin films grown by atomic layer deposition (ALD). In particular we show that the doping is uniform, with
high iron concentration and no segregation, and that zirconia is in the
tetragonal/cubic structure. Thus theoretically we focus our attention on the 
tetragonal structure of zirconia with substitutional iron doping uniformly distributed
in the sample.
In Sec.~\ref{sec:magnetic_properties}
we study the magnetization of the system and how it is influenced by defects, i.e.
oxygen vacancies ($V^{^{\bullet\bullet}}_O$),
comparing our results with the IBM and the CMM.
Indeed in a recent work we showed that iron doping induces $V^{^{\bullet\bullet}}_O$,
with a ratio $y_{V^{^{\bullet\bullet}}_O/Fe}=0.5$, for charge compensation, and
that $ZrO_2$:$Fe$ films growth by ALD
presents a ratio close to one half~\cite{Sangalli2012}.
We finally investigate the magnetization of the films growth by ALD.
From the magnetization at saturation we extract the magnetic moment per atom which
is discussed in view of the results from theoretical simulations.

\section{Structural Properties of $\mathbf{ZrO_2}$:$\mathbf{Fe}$}
\label{sec:structural_properties}

In order to describe the effect of iron doping in zirconia,
$ZrO_2$:$Fe$ thin films were grown on $Si/SiO_2$ substrates in a flow--type hot wall
atomic layer deposition reactor (ASM F120) starting from $\beta$--diketonates metalorganic
precursors. Ozone was used as oxidizing gas in the reaction process.
The $Fe$ concentration in $ZrO_2$:$Fe$ films can be tuned tailoring the $Zr/Fe$
precursors pulsing ratio. In the present work however we focused our attention
on the high doping regime keeping the pulsing ratio fixed. The growth temperature was
maintained at $350^\circ$C. After the deposition the films were annealed at $600^\circ$C
in $N_2$ flux for 60s to study their thermal stability. Further details on the samples
preparation can be found in Ref.~\cite{Lamperti2012}.

\begin{figure}
\begin{center}
\resizebox{0.49\columnwidth}{!}{\includegraphics{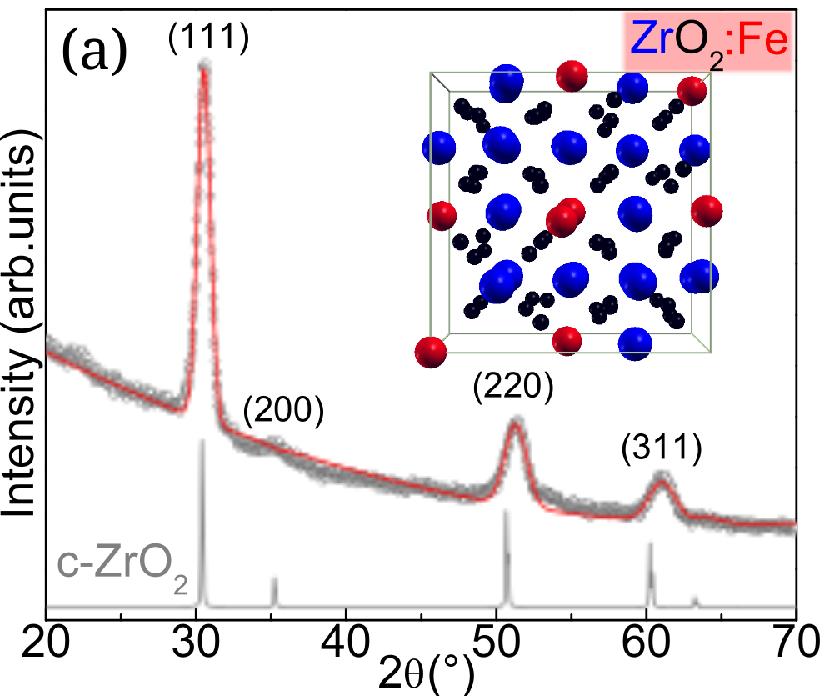} }
\resizebox{0.49\columnwidth}{!}{\includegraphics{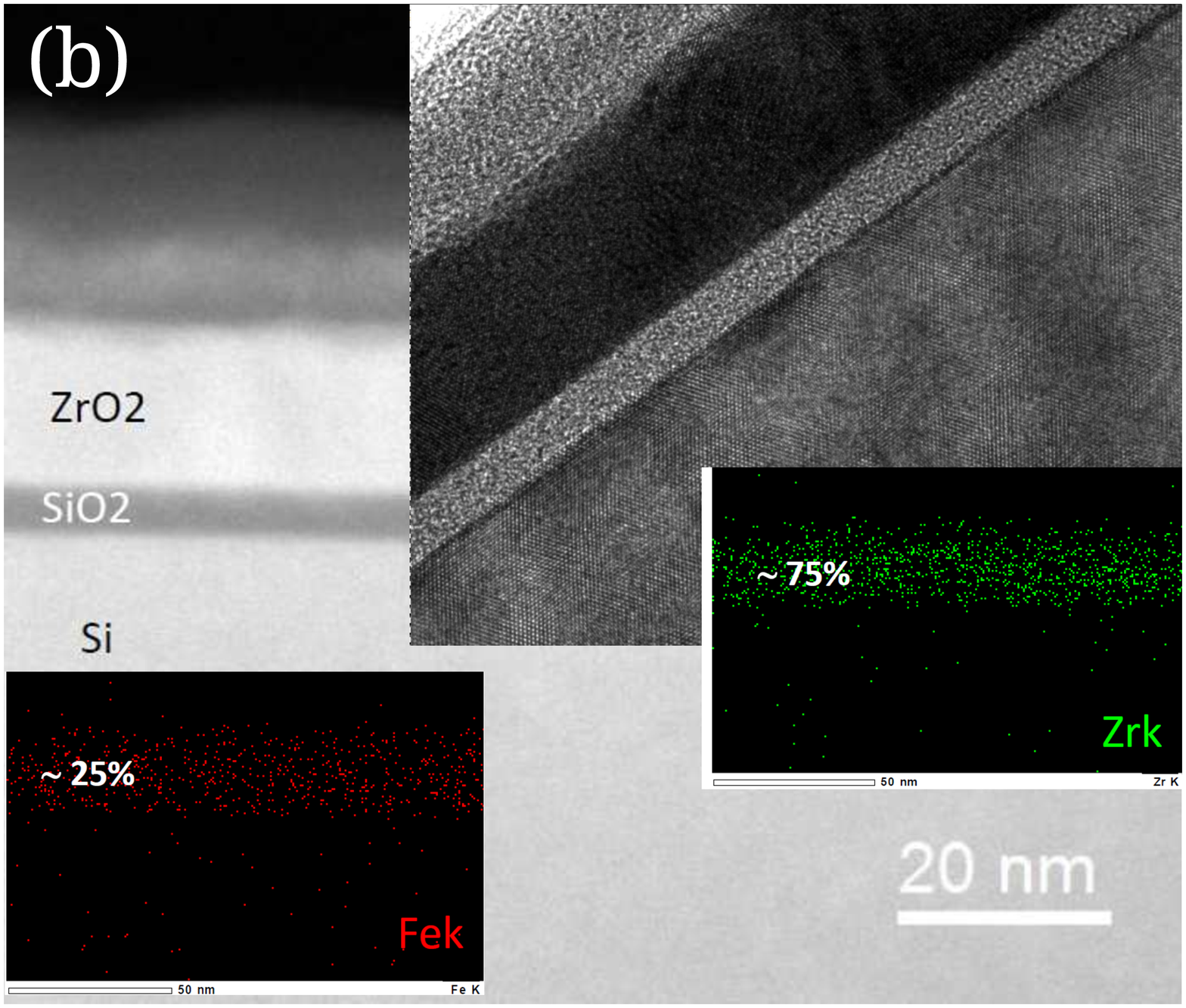} }
\caption{(colors online) Structural properties of iron doped zirconia.
Panel $(a)$: XRD patterns show that the film presents the cubic/tetragonal phase~\cite{XRD_structure}.
A representation of the fully relaxed theoretical structure is also pictured
with $Fe$ atoms in red, $Zr$ atoms in blue and oxygens in black (we used the Xcrysden software~\cite{Xcrysden}).
Panel $(b)$: HAADF image and HREM inset showing the layered geometry of the sample.
The EDX chemical maps show a uniform doping distribution with no segregation.
}
\label{fig:TEM+XRD}
\end{center}
\end{figure}

Film crystallinity was checked by X--ray diffraction (XRD) at fixed grazing incidence angle $\omega=1^\circ$
and using $Cu$ $K_\alpha$ ($\lambda = 0.154$ nm) monochromated and collimated X--ray beam
(Italstructure XRD 3000, details on the measurements can be found in Ref.~\cite{Lamperti2011}).
In the present work all measures shown are from the same film which we chose as representative
of the high--doping concentration samples. 
Fig~\ref{fig:TEM+XRD}$(a)$ shows that the films present a cubic/tetragonal crystalline structure
with an estimated cell parameter, assuming a cubic cell, $a=5.024$ \AA\ with a contraction of
about $(a_{exp}-a_{exp}^0)/a_{exp}^0\approx -1.0\%$; here $a_{exp}^0=5.074$ is the lattice parameter
measured for undoped $ZrO_2$ films.
Theoretically we found, at $x=25\%$, $(a_{theo}-a_{theo}^0)/a_{theo}^0\approx -0.4\%$, with
$a_{theo}^0=5.11$, the lattice parameter computed for $ZrO_2$,
while the tetragonal deformation reduces from $3.05\%$ to  $<1.\%$.

The samples were then characterized by transmission electron microscopy (TEM),
performed by a JEOL 2200FS microscope equipped with
a high--angle annular--dark--field (HAADF) detector, in--column energy filter and
energy dispersive X--ray (EDX) spectrometer.
The layered geometry of the samples is shown in fig.~\ref{fig:TEM+XRD}$(b)$.
The $Fe$ atomic concentration measured by EDX is $25\%$, in agreement within few percent
to the estimation done by X--ray photo--emission (XPS).
The EDX maps, taken in different regions of the samples, showed that $Fe$ is uniformly distributed
across the film and no clusters or segregation at the grain boundaries have been found
(see Fig.~\ref{fig:TEM+XRD}$(b)$, insets).

Thus theoretically we computed, from first--principles,
the ground state of the the tetragonal phase of $ZrO_2$:$Fe$ with uniform iron doping distribution.
We used the PWSCF~\cite{QuantumEspresso} package considering a super--cell with 96 atoms; for all systems
the atomic positions are fully relaxed~\cite{Footnote1}. The ground state was computed with 
the generalized gradient approximation~\cite{Perdew1996} (GGA) to the density functional theory (DFT)
scheme~\cite{Hohenberg1964,Kohn1965} with ultra--soft pseudo--potentials~\cite{Vanderbilt1990,Rappe1990}.
We have recently shown that iron, in the zirconia host lattice, behave like Yttrium ($Y$)
which is among the most studied dopant of this oxide: it replaces zirconium atoms in the $ZrO_2$
lattice inducing $V^{^{\bullet\bullet}}_O$ for charge compensation, with a ratio $y_{V^{^{\bullet\bullet}}_O/Fe}=0.5$,
and stabilizes the tetragonal phase above $x_{Fe}\approx11\%$~\cite{Sangalli2011,Sangalli2012}.
Calculations have been done at different doping concentrations, $x_{Fe}=6.24,\ 12.5,\ 25.0\%$ and
different $V^{^{\bullet\bullet}}_O$ concentration, considering the $V^{^{\bullet\bullet}}_O$ to $Fe$
ratios $y_{V^{^{\bullet\bullet}}_O/Fe}=0,\ 0.5,\ 1$,
focusing our attention at the high doping concentration limit, 
i.e. the concentration measured experimentally.
In the inset of Fig.~\ref{fig:TEM+XRD}$(b)$ a relaxed structure at $y_{V^{^{\bullet\bullet}}_O/Fe}=0.5$
and $x_{Fe}=25\%$, which is the expected structure in our films~\cite{Sangalli2012}, is shown.

In the next section we discuss in details how the magnetic properties of the system
and its electronic structure are correlated. In particular the density of states (DOS) at the Fermi
level can crucially affect the exchange mechanisms in the system. The result presented here
are obtained within the GGA approximation which is known however to suffer of the self--interaction
problem and thus to de--localize too much the $d$ electrons. Thus the so called Hubbard $U$ parameter
(GG+U scheme) is often used. Indeed in Ref.~\cite{Sangalli2012} we have discussed
how this correction modify the DOS of the system. 
However we have also shown that the $U=0.0\ eV$ choice better reproduces the experimentally measured
band structure. Accordingly in the present manuscript we discuss the magnetic properties in this 
configuration, i.e. within the simple GGA scheme, pointing out how the Hubbard $U$ correction
would eventually modify the conclusions of our approach.

\section{Magnetic Properties of $ZrO_2$:$Fe$}
\label{sec:magnetic_properties}

\begin{figure}
\begin{center}
\resizebox{0.93\columnwidth}{!}{\includegraphics{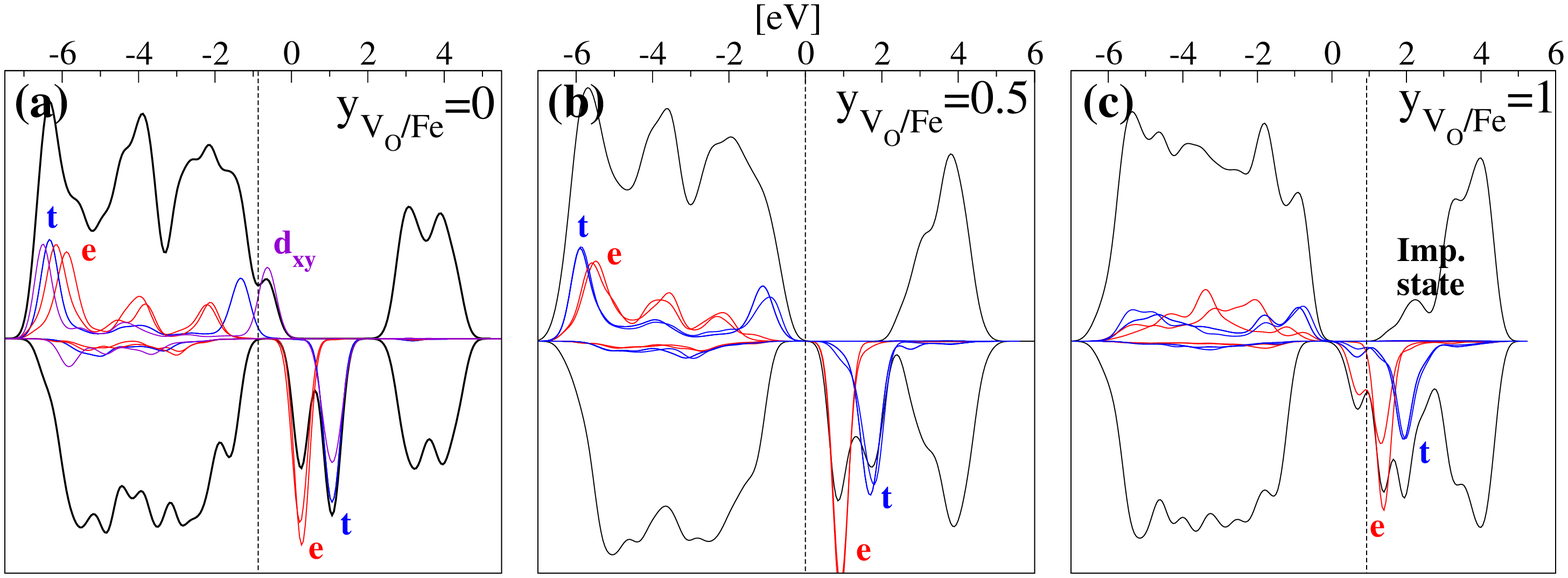}}
\resizebox{0.31\columnwidth}{!}{\includegraphics{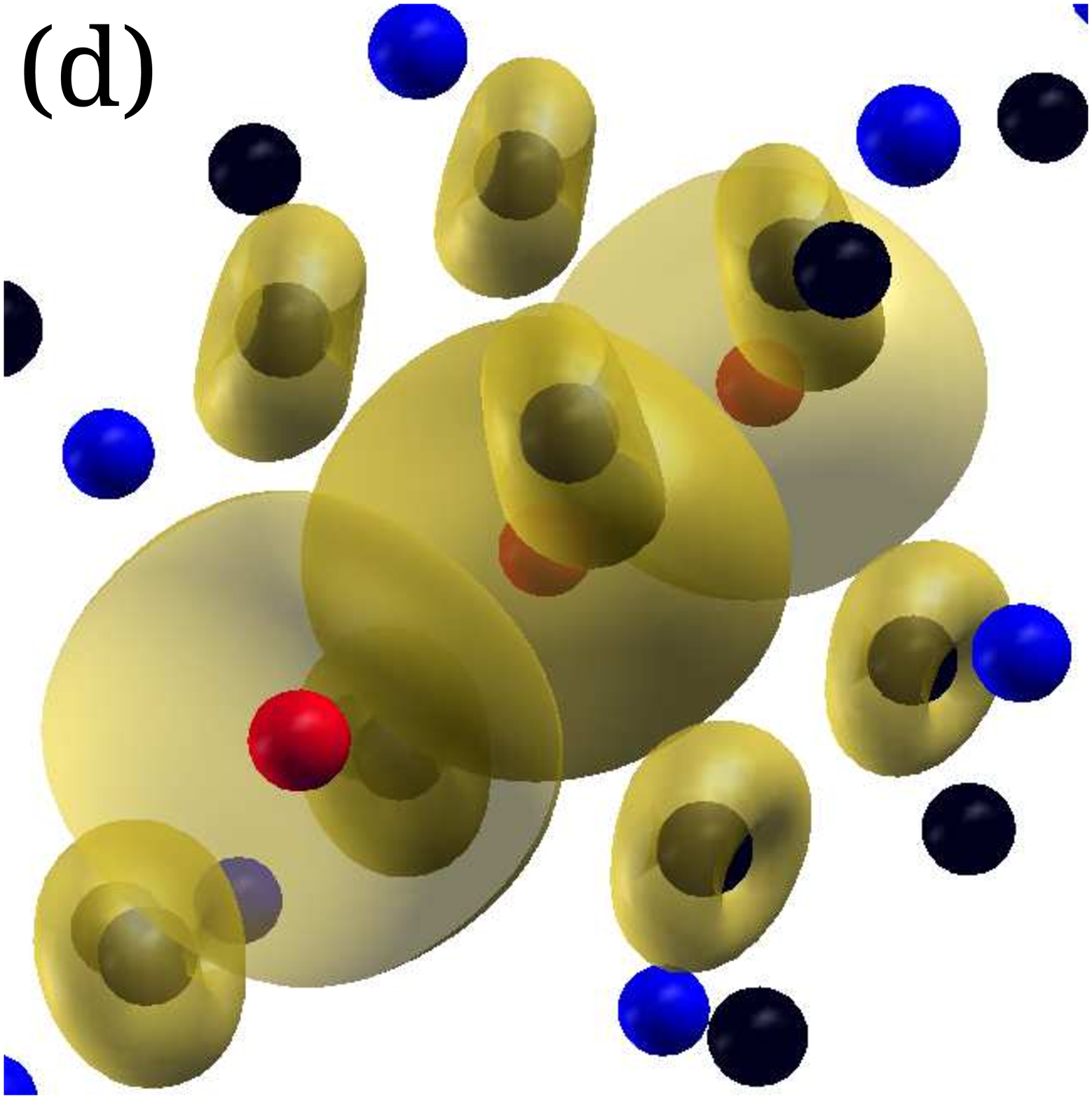}}
\resizebox{0.31\columnwidth}{!}{\includegraphics{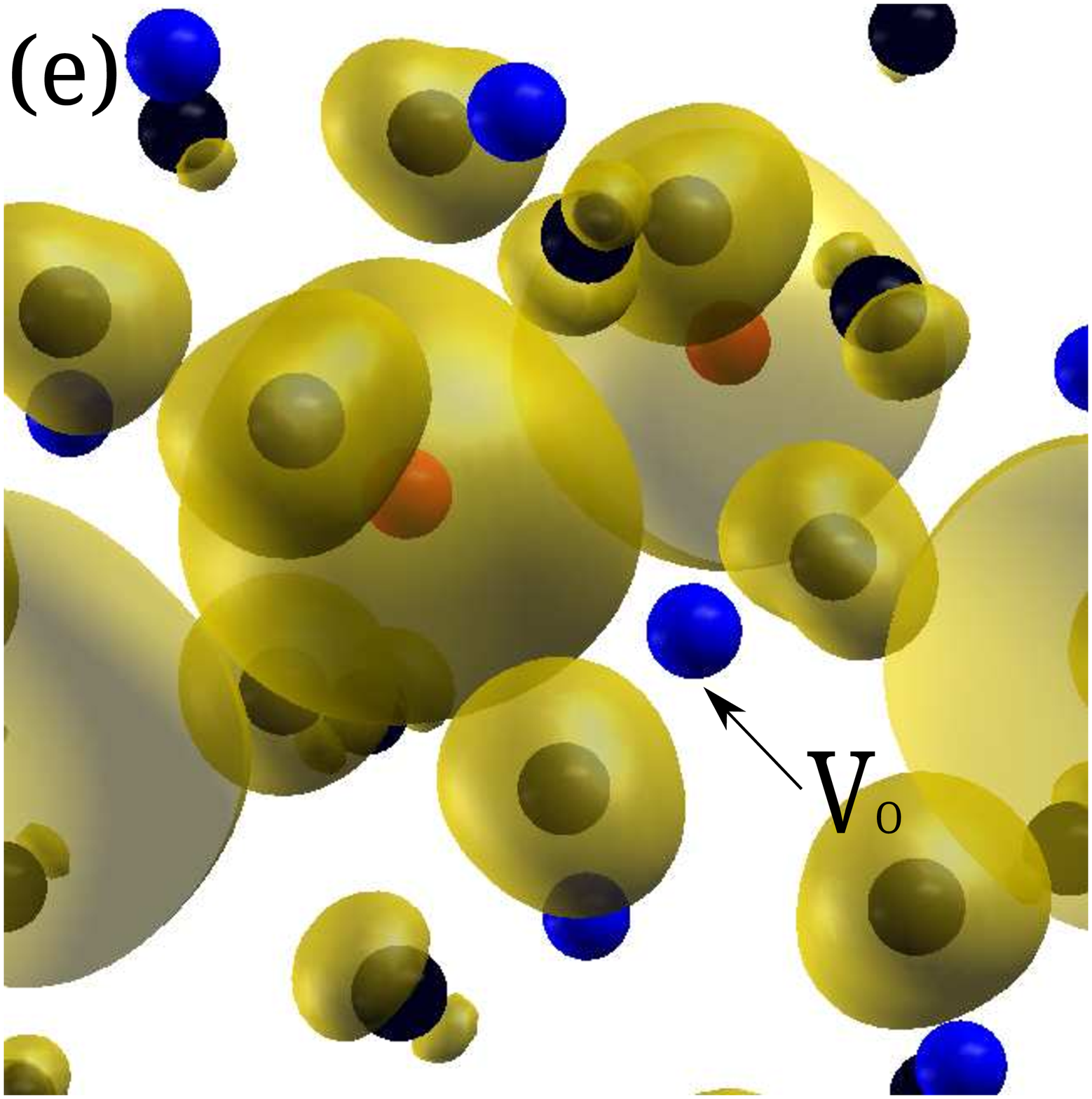} }
\resizebox{0.31\columnwidth}{!}{\includegraphics{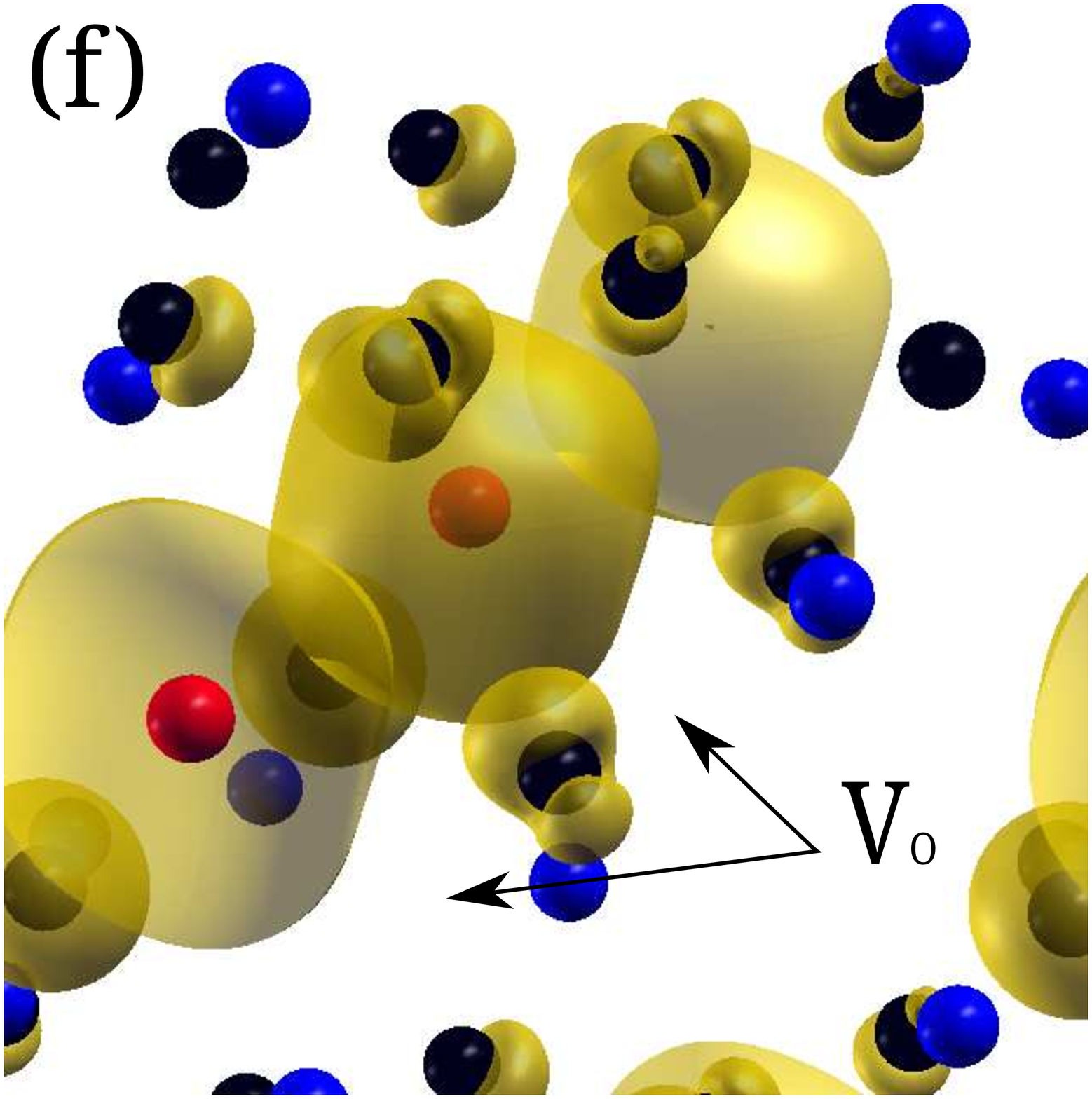} }
\caption{(colors online) Theoretical density of states (panels $a$-$c$) and magnetization density
isosurfaces at $m=0.01\ a.u.$
(panels $d$-$f$) for the ferromagnetic configuration at $y_{V^{^{\bullet\bullet}}_O/Fe}=0,\ 0.5,\ 1$.
The density of states projected on the $d$ atomic orbitals
is also shown. The crystal field splitting is visible with the $e$ doublet, $d_{z^2}$ and $d_{x^2-y^2}$ in blue,
at higher (lower) energy in the majority (minority) spin channel respect to the $t$ triplet,
$d_{xy}$, $d_{xz}$ and $d_{yz}$ in red.
In panels $d$-$f$ $Fe$ atoms are represented in red, $Zr$ atoms in blue and oxygens in black.}
\label{fig:Magn_theo}
\end{center}
\end{figure}

\begin{figure}
\begin{center}
\begin{multicols}{2}{
\resizebox{0.99\columnwidth}{!}{\includegraphics{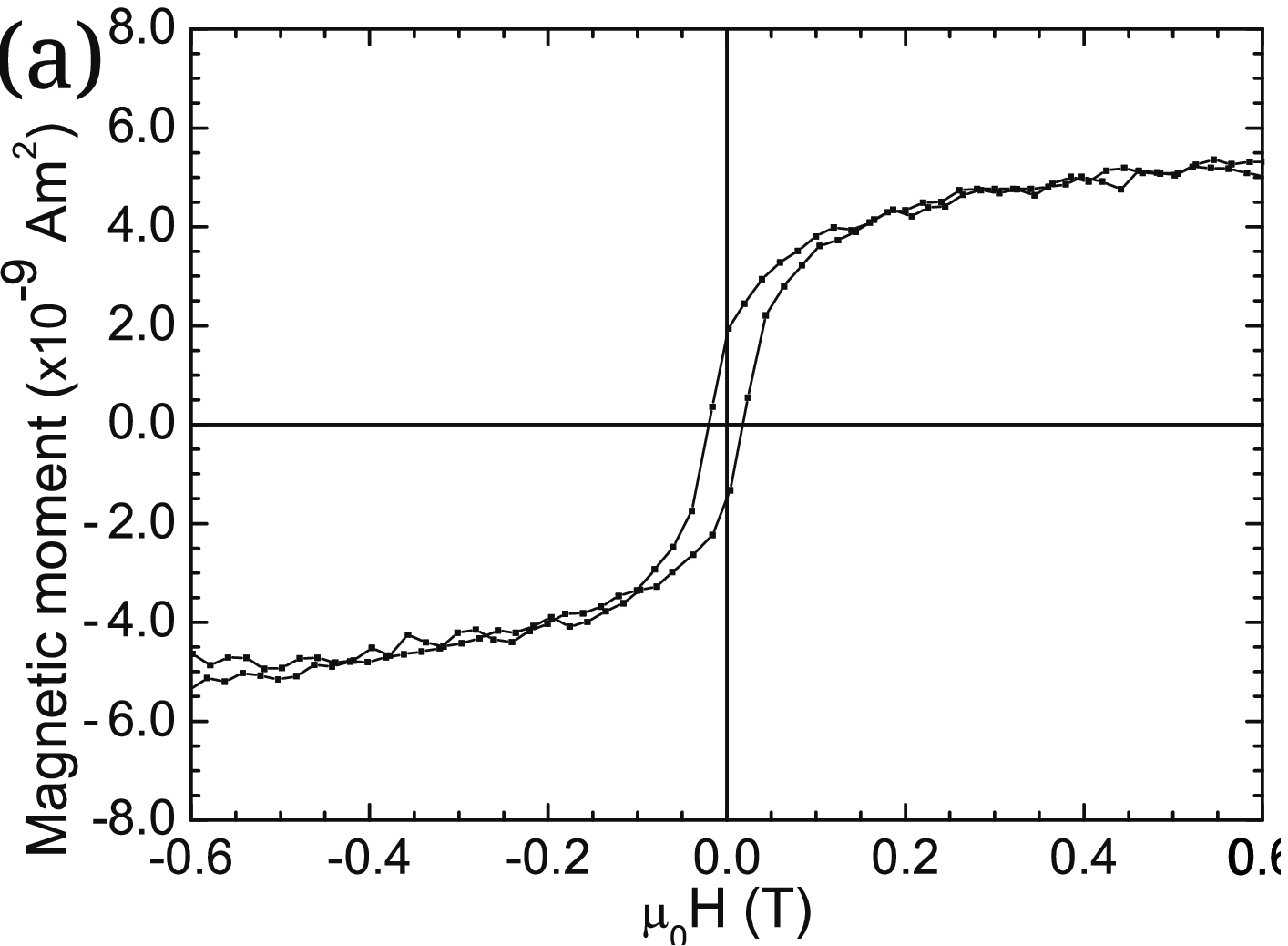} } \\
\resizebox{0.99\columnwidth}{!}{\includegraphics{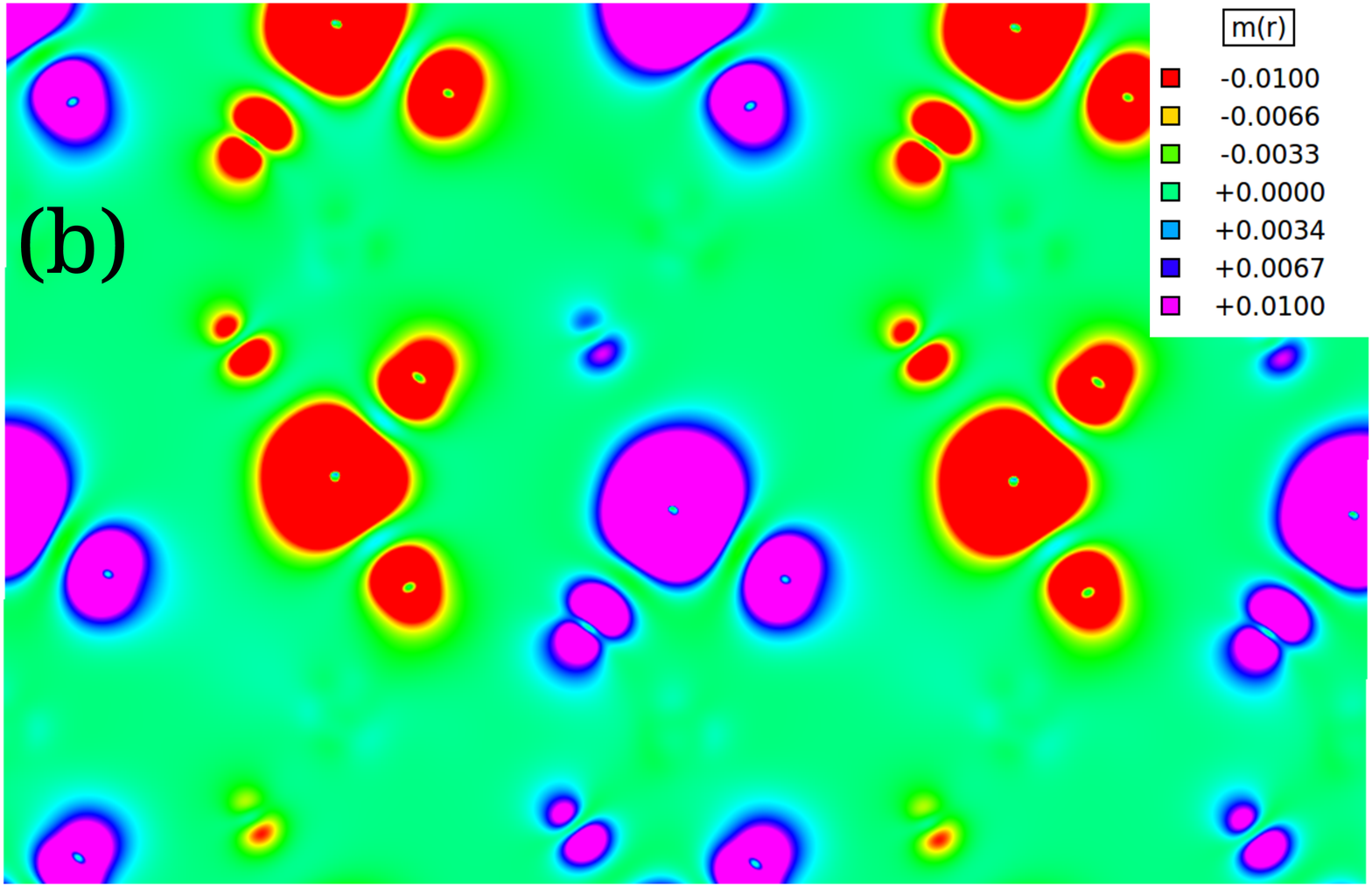} } \\
\resizebox{0.99\columnwidth}{!}{\includegraphics{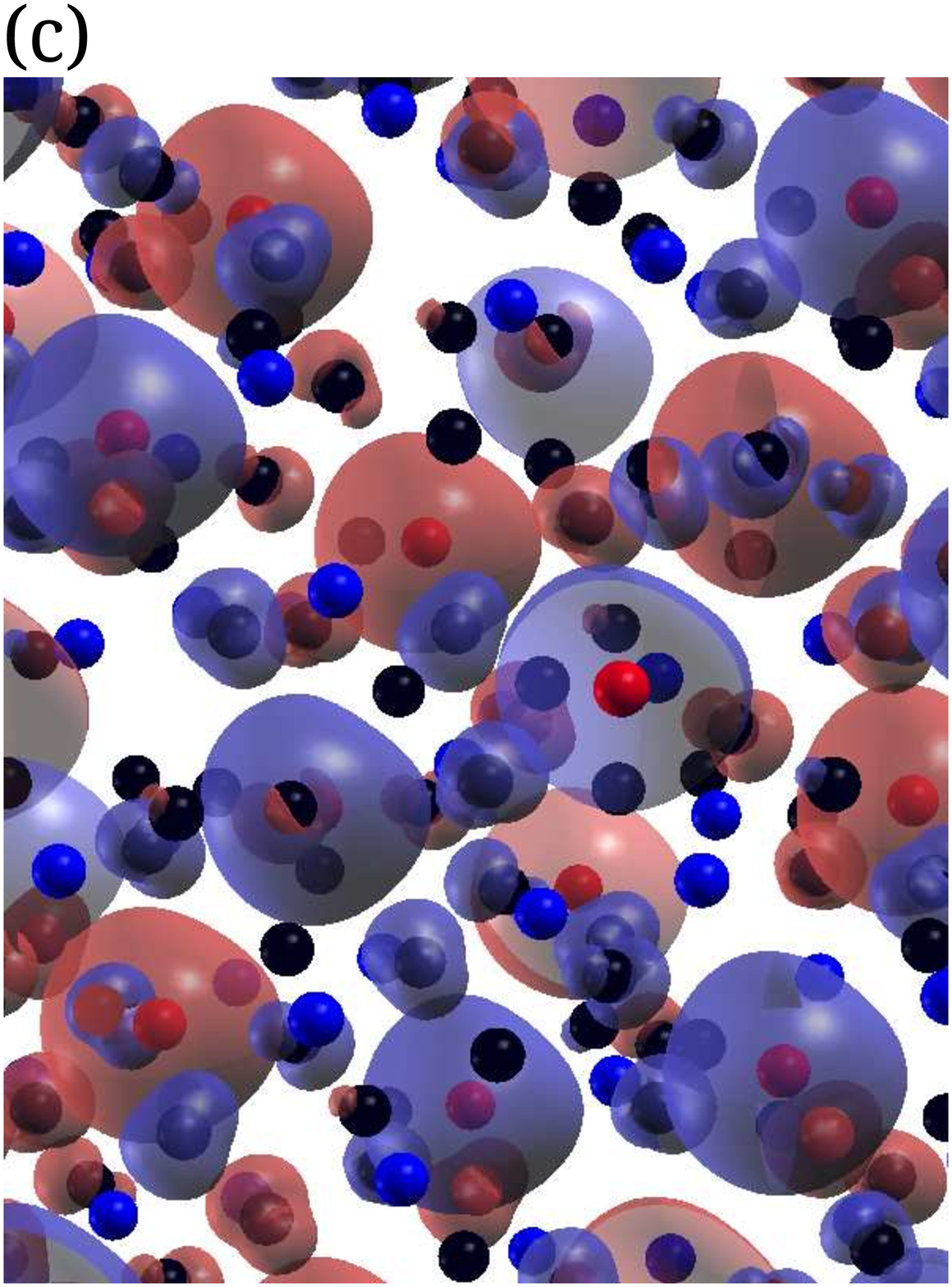} }
}
\end{multicols}
\caption{(colors online) Panel $a$: magnetization hysteresis loop measured by
Alternating gradient force magnetometer at room temperature.
The resulting magnetization at saturation can be interpreted in terms of a anti--ferromagnetic
structure if $y_{V^{^{\bullet\bullet}}_O/Fe}$ deviates from 0.5 of about $10\%$. The magnetization density
for the anti--FM configuration at $y_{V^{^{\bullet\bullet}}_O/Fe}=0.5$ is plotted in panels $(b)$-$(c)$.
Panel $b$: 2D contour plot of the magnetization density along the plane $(1\ $-$1\ 0)$.
Panel $c$: Positive (red) and negative (blue) isosurfaces of the magnetization density at $m=\pm 0.02$.
 }
\label{fig:Magn_exp}
\end{center}
\end{figure}

Starting from the fully relaxed structures considered at $x_{Fe}=25\%$ we computed the
local magnetization of the system. In Fig.~\ref{fig:Magn_theo} the magnetization density,
i.e. the difference between the spin--majority and the spin--minority density is represented
for $y_{V^{^{\bullet\bullet}}_O/Fe}=0,\ 0.5,\ 1$. In order to compare the
present results with the IBM also the total DOS are plotted.
The FM configuration is considered for better clarity as it helps to distinguish
between the minority and the majority spin channel. However all the conclusion we will draw
in the following also hold for the anti--FM configuration (see Fig.~\ref{fig:Magn_exp}$(b$-$c)$
for a comparison of the anti--FM magnetization density).

At $y_{V^{^{\bullet\bullet}}_O/Fe}=0$, the system shows holes in the valence band (Fig.~\ref{fig:Magn_theo}$(a)$)
because iron act as an  acceptor;
the projected DOS shows that these are localized on the $Fe(d_{xy})$ level.
As a consequence the majority--spin $d$ band is not completely filled, the magnetic
moment per iron atom is $4\ \mu_B$ and iron is forced in a $+4$ oxidation state.
This also breaks the spherical symmetry of the system and
the shape of the magnetization density close to the oxygens is not isotropic, with
a peculiar doughnut shape oriented in the $xy$--plane.

Creating $V^{^{\bullet\bullet}}_O$, i.e. increasing $y_{V^{^{\bullet\bullet}}_O/Fe}>0$, electrons
are released in the system.
These fill the empty $d_{xy}$ levels with a charge--transfer mechanism.
At $y_{V^{^{\bullet\bullet}}_O/Fe}=0.5$ the majority $d$ band is completely filled, the magnetic moment per
iron atom is $5\ \mu_B$ and the magnetization has a spherical shape. The system
turns into a charge--transfer semi--conductor (Fig.~\ref{fig:Magn_theo}$(b)$).
At $y_{V^{^{\bullet\bullet}}_O/Fe}=1$ electrons fill the minority $d$ levels and the magnetic moment per
iron atom is reduced to $4\ \mu_B$. However the minority $d$ electron does not participate in
the bonds, thus the anisotropy is only weakly transferred to the $p$ orbitals of oxygen
(Fig.~\ref{fig:Magn_theo}$(c)$).
In all configurations the magnetization is mainly located around the $Fe$ atoms
and it is in part transferred to the oxygen next nearest neighbor suggesting a short range
magnetic interaction (Fig.~\ref{fig:Magn_theo}$(d$-$f)$). 

It is worth to compare the DOS of our system with the IBM. A crucial assumption of the 
IBM is that the presence of $V^{^{\bullet\bullet}}_O$ creates impurity states with poorly localized
electrons. These, already at low doping concentration, would overlap to create an impurity
band that can mediate the FM interaction among nearby iron atoms.
Indeed $V^{^{\bullet\bullet}}_O$ are a common defect in pure $ZrO_2$ films where they create impurity states
close to the conduction band.
However in the case for $ZrO_2$:$Fe$ the situation is different, as shown in
Fig.~\ref{fig:Magn_theo}$(a$-$c)$.
For low $V^{^{\bullet\bullet}}_O$ concentration, i.e. $y_{V^{^{\bullet\bullet}}_O/Fe}\leq 0.5$
no impurity states are associated
to $V^{^{\bullet\bullet}}_O$ while for $x>0.5$ impurity states appear but higher
in energy than the empty spin minority $d$-levels. Thus the extra electrons
are trapped in the $Fe(d)$ levels and the impurity states remain empty.
More in general, the majority of magnetic transition metals (TMs),
i.e. $Fe$, $Co$, $Ni$, $Mn$, and $Cr$, have $+2$ or $+3$ as most stable oxidation state and we expect
a similar picture for all TMs doped $XO_2$ oxides, at least for $y_{V^{^{\bullet\bullet}}_O/Fe}\leq0.5\ (1.0)$, in
the case of $+3\ (+2)$ oxidation state.
Thus, according to our results, the IBM cannot be invoked to describe the magnetic ground--state
for $ZrO_2$:$Fe$ and for $XO_2$:$TM$ in general.

The situation depicted is much closer to the CMM proposed for $Ga_{1-x}Mn_xAs$.
In $Ga_{1-x}Mn_xAs$ $Mn$ acts as an acceptor which compensates the 
anti--site defects commonly present in $GaAs$ giving a charge--transfer semiconductor;
in $ZrO_2$:$Fe$ $Fe$ acts as an acceptor compensating the $V^{^{\bullet\bullet}}_O$. The main difference
between the two systems is the nature of the host lattice, a covalent semiconductor
the former and a polar oxide the latter. In $Ga_{1-x}Mn_xAs$, when
$x_{Mn}$ exceeds the anti site defect concentration,
the system behaves like a metal with holes in the valence band.
In $ZrO2:Fe$ instead, at $y_{V^{^{\bullet\bullet}}_O/Fe} < 0.5$,
the holes are localized onto the $d_{xy}$ states
(see Fig.~\ref{fig:Magn_theo}$(b)$); conduction could be possibly obtained only
at high doping concentration with a hopping like mechanism.

The main correction to the GGA electronic structure due to the Hubbard $U$ term is to shift 
down in energy the $d$ orbitals. In particular at $y_{V^{^{\bullet\bullet}}_O/Fe}=0$
this tends to de--localize the hole in the valence band from the iron $d$ orbitals
to the oxygen $p$ levels. At $y_{V^{^{\bullet\bullet}}_O/Fe}=0.5$ no
other qualitative change in the DOS can be observed.
Finally at $y_{V^{^{\bullet\bullet}}_O/Fe}=1$ the Hubbard $U$ correction opens a gap
between the occupied and the empty $d$ orbitals in the minority spin channel.
However all the consideration on the IBM and the CMM still holds. The de--localization of
the holes for $y_{V^{^{\bullet\bullet}}_O/Fe}<0.5$ could possibly make the system
more similar to standard DMS as $Ga_{1-x}Mn_xAs$. However we remind that in the case
of $ZrO2:Fe$ the Hubbard $U$ correction doe not improve the description of the electronic
properties of the system over standard GGA~\cite{Sangalli2012}. Also the de--localization
of holes in oxides is under debate~\cite{Varley2012}.

We then focus our attention on the strength of the magnetic interaction in our system.
This in principle can be understood computing the energy difference between the FM
and the PM phase. The latter however can be hardly described within a periodic
code, as the description of random magnetic moment orientations would require
huge super--cells with a non--collinear description of the wave--functions. 
The energy difference per iron atom between the FM and the anti--FM
configuration however can be used as a reasonable approximation~\cite{Sato2010}.
We found that at lower doping, $x_{Fe}=6.25,\ 12.5\%$,
the energy difference is very low with $\Delta E/k_B$ of the
order of few kelvin~\cite{footnote1}.
The anti--FM configuration is slightly favored at $y_{V^{^{\bullet\bullet}}_O/Fe}=0.5$, while the FM
one is slightly favored at $y_{V^{^{\bullet\bullet}}_O/Fe}=0$.
Instead at $x_{Fe}=25\%$ in both cases the anti--FM configuration is favored
with $\Delta E/k_B\approx 150 K$ for $y_{V^{^{\bullet\bullet}}_O/Fe}=0.5$ and
$\Delta E/k_B\approx 20 K$ for $y_{V^{^{\bullet\bullet}}_O/Fe}=0$.
This suggests a short range interaction
which becomes relevant at $x_{Fe}>x_P$, with $x_P$ the percolation threshold,
i.e. an anti--FM super--exchange mechanism which is dominant 
in the $y_{V^{^{\bullet\bullet}}_O/Fe}=0.5$ case.
For the $y_{V^{^{\bullet\bullet}}_O/Fe}=0$ configuration however this super--exchange mechanism
appears to be in competition with a FM interaction which could be explained
in terms of the CMM model,
though with a weaker effect due to the low mobility of the holes.
This results in a weakly FM interaction at low doping and a weakly anti--FM
interaction at higher doping. In this case the prediction of the GGA+U approach are
substantially different. Indeed the U correction,
de--localizing the holes,
strongly enhances the CMM mechanism and indeed the systems turns out to be ferromagnetic at 
any doping concentration with $\Delta E/k_B\approx 150 K$ at $x_{Fe}=25\%$.

Experimentally the magnetic properties of sample with  $x_{Fe}\approx25\%$ were studied by means of alternated
gradient force magnetometry (AGFM). Magnetization measurements, performed at room temperature by applying the
magnetic field parallel to the film plane, show a clear hysteresis loop characterized by a coercive
field $\mu_0H_c\approx 0.03\ T$ (Fig. \ref{fig:Magn_exp}$(a)$).
From the saturation magnetization value a magnetic moment per $Fe$ atom $m \approx 0.6\mu_B$
can be extrapolated. This is consistent with the values reported in the literature but is one
tenth of the theoretically predicted value, $m \approx 5\mu_B$. Moreover theoretically we found that
the anti--FM configuration is the most stable at least for $y_{V^{^{\bullet\bullet}}_O/Fe} = 0.5$ which
is the expected situation in our films~\cite{Sangalli2012}.
In Fig.~\ref{fig:Magn_exp}$(b$-$c)$ we plotted the magnetization density in the
anti--FM configuration for $y_{V^{^{\bullet\bullet}}_O/Fe}=0.5$. As in the case of FM ground state,
the magnetization is mainly located on the $Fe$ atoms and the next nearest neighbor oxygens.

The discrepancy between the theoretically and the experimental results could be possibly
explained supposing that experimentally $y_{V^{^{\bullet\bullet}}_O/Fe}=0.5$ were not fulfilled
by few percent of the $Fe$ atoms, which thus would be in the $Fe^{4+}/Fe^{2+}$ oxidation state.
If we assume that these ions would couple with FM interaction, with a CMM like mechanism,
the net results would be a weak magnetic system with low magnetic moment per atom
Also in this case however the energy difference between the FM and the anti--FM configuration
is too low to explain the experimental result. A possible
mechanism to explain the room temperature magnetism is suggested by the LDA+$U$ approach where
the de--localization of the holes strongly enhance the CMM mechanism.
However further investigations are needed, both experimentally and
theoretically, to clarify this point.

\section{Conclusions}
\label{sec:conclusions}
In conclusion we have studied both theoretically and experimentally thin films of iron doped
$ZrO_2$. Experimentally we have shown that iron distributes uniformly at high doping concentration
with no cluster formation and that the system is magnetized at room temperature.
With first principles simulations we have discussed the possible magnetization mechanisms
comparing our results with other model proposed in the literature.
We showed that the impurity band model cannot be invoked in the case of iron doped zirconia
while the carrier mediated model could be possibly considered for
uncompensated oxygen vacancies concentration (i.e. $y_{V^{^{\bullet\bullet}}_O/Fe}\neq0.5$).
The standard super--exchange mechanism however appears to be dominant 
for $y_{V^{^{\bullet\bullet}}_O/Fe}=0.5$.

From magnetization measurements of highly doped samples we found a low saturation magnetization value
corresponding to low magnetic moment per Fe atom.
We tried to interpret the experimental results in view of our theoretical findings.
However the difference between theory and experiments suggest that other effects need
to be taken into account for a correct description of the magnetic properties of the system.
The role of the Hubbard $U$ correction together with the deviation from the theoretically
ideal $y_{V^{^{\bullet\bullet}}_O/Fe}=0.5$ configuration have been proposed as possible
candidates.

\section{Acknowledgements}
This work was funded by the Cariplo Foundation through the OSEA project (n.~2009-2552).
Helpful discussions with L. Lazzarini are gratefully acknowledged.
D.S. and A.D. would like to acknowledge G. Onida and the ETSF Milan node
for the opportunity of running simulations on the ``etsfmi cluster'',
and P. Salvestrini for technical support on the cluster.
We also acknowledge computational resources provided by the Consorzio Interuniversitario
per le Applicazioni di Supercalcolo Per Universit\`a e Ricerca (CASPUR) within the project MOSE.


\begin{thebibliography}{}
\bibitem{Footnote1} We use a cut---off of 35 Ry for the wave--functions and 400 Ry for the augmentation density;
a Monkhorst--Pack grid 2x2x2 for the Brillouin zone.
\bibitem{footnote1} The ferromagnetic structure at $x_{Fe}=12.5\%$ is frustrated. Considering the closest $Fe$-$Fe$
atoms we have in total, within in the supercell, 6 anti-ferromagnetic ``connections'' and 2 ferromagnetic connections.
\end{thebibliography}
\end{document}